\documentclass[%
 reprint, 
 amsmath,amssymb,
 aip,
floatfix,
rsi,
]{revtex4-1}
\usepackage{graphicx}
\usepackage{dcolumn}
\usepackage{bm}

\usepackage{nicefrac}
\usepackage{url}

\newcommand{\rfig}[1]{Figure~\ref{figure_#1}}

\newcommand{\basicfig}[3]{%
      \begin{figure}[tbp]%
      \centering%
      	  #1
      \caption{#3}
      \label{figure_#2}
      \end{figure}
}

\newcommand{\boxfig}[3]{\basicfig{\mbox{#1}}{#2}{#3}}

\newcommand{\nfig}[4][]{ 
      \boxfig{\includegraphics[#1]{#2}}{#3}{#4}
}
\newcommand{\fig}[3][]{\nfig[#1]{#2}{#2}{#3}}

\begin{document}

\preprint{APS/123-QED}

\title{A one-piece 3D printed flexure translation stage for open-source microscopy}

\author{James P. Sharkey}
\author{Darryl C. W. Foo}
\affiliation{%
 Nanophotonics Centre, Department of Physics, Cavendish Laboratory, University of Cambridge, CB3 0HE, UK
}%

\author{Alexandre Kabla}
\affiliation{
 Department of Engineering, University of Cambridge, CB2 1PZ, UK
}%

\author{Jeremy J. Baumberg}
 \homepage{http://www.np.phy.cam.ac.uk/}
\author{Richard W. Bowman}%
\email{richard.bowman@cantab.net}
 \altaffiliation[Also at ]{Queens' College, Cambridge.}
\affiliation{%
 Nanophotonics Centre, Department of Physics, Cavendish Laboratory, University of Cambridge, CB3 0HE, UK
}%

\date{\today}

\begin{abstract}
Open source hardware has the potential to revolutionise the way we build scientific instruments; with the advent of readily-available 3D printers, mechanical designs can now be shared, improved and replicated faster and more easily than ever before.  However, printed parts are typically plastic and often perform poorly compared to traditionally machined mechanisms.  We have overcome many of the limitations of 3D printed mechanisms by exploiting the compliance of the plastic to produce a monolithic 3D printed flexure translation stage, capable of sub-micron-scale motion over a range of $8\times8\times4\,$mm.  This requires minimal post-print clean-up, and can be automated with readily-available stepper motors.  The resulting plastic composite structure is very stiff and exhibits remarkably low drift, moving less than $20\,\mu$m over the course of a week, without temperature stabilisation.  This enables us to construct a miniature microscope with excellent mechanical stability, perfect for timelapse measurements \textit{in situ} in an incubator or fume hood.  The ease of manufacture lends itself to use in containment facilities where disposability is advantageous, and to experiments requiring many microscopes in parallel.  High performance mechanisms based on printed flexures need not be limited to microscopy, and we anticipate their use in other devices both within the laboratory and beyond.

\end{abstract}

\pacs{Valid PACS appear here}
\maketitle

\section{Introduction}
The need to precisely position samples, probes, and other items is a ubiquitous challenge when designing apparatus; good mechanical design is essential for most scientific experiments.  Often, the constraints of a given experiment mean some level of customisation is required, which entails difficult, time-consuming mechanical design and production of one-off parts.  For example, tightly-integrated mechanical assemblies are often preferable to stacking multiple commercially available translation stages due to their lower drift, higher stiffness, and ability to fit around mechanical constraints -- but often difficulty of manufacture means the latter option is taken, to the detriment of experimental performance.

3D printing has recently emerged as a readily-available technology, thanks largely to the open-source RepRap project\cite{Jones:2011}, where the designs for a printer (made using standard components and printed plastic parts) were shared online for the community to use and improve.  Automated desktop machines that form plastic parts by extruding a filament of molten plastic (``fused filament fabrication'') are fast becoming a standard item of equipment in many laboratories and workshops.  The ability to conveniently produce accurate parts from digital designs has led to an explosion of interest in other devices that can be produced and improved in this way.  In a scientific context, access to the design of an instrument -- such as the OpenSPIM microscope \cite{Pitrone:2013} -- allows a deeper understanding of its performance and limitations, and facilitates customisations and improvements.  These modifications can then be easily shared, enabling open source instruments to improve rapidly in performance and versatility.  The primary benefit of open source scientific hardware, as with software, is not lower cost; it is the ability to investigate, verify and improve the method\cite{Pearce:2014,Ince:2012,Prlic:2012,openlabtools}.

A wide variety of printable designs is freely available online, but the low stiffness and poor surface finish of printed parts compared to machined ones mean that high-performance mechanisms are rarely achieved by using the same designs that are machined from metal.  Furthermore, many designs involving motion have complex build instructions with many non-printed parts that must be obtained and assembled.  Motion control is a good example of this; sliding dovetail stages are a staple mechanical component for displacements ranging from millimetres to centimetres, but the roughness and poor size tolerance of printed parts makes it prohibitively difficult to print reliable linear stages.  However, translation stages based on flexures do not have these requirements; their main drawback is that, when machined from metal, the range of motion is usually limited by material stiffness.  We have implemented flexures in a 3D printed structure, using a design optimised for printed plastic rather than metal.  The result is a highly stable and precise translation stage with a greater range than equivalently-sized metal flexures.  The main mechanism of our stage is a monolithic printed part, requiring minimal post-print assembly and very few additional items.  It is a parametric design, implemented in OpenSCAD\cite{OpenSCAD,*CGAL,*Slic3r}, making it simple to alter the size or design of the stage, and can be printed on the vast majority of currently available printers.

Microscopy is a technique where precise positioning is critically important; the sample to be observed must be held stably in the focal plane, and translated to place features of interest in the field of view of the objective lens.  We have implemented a simple optical microscope based around our printed translation stage (\rfig{photograph}) to allow us to quantify its mechanical performance in a realistic situation.  This allows us to measure its stability over a range of timescales, and to demonstrate the precision with which it can position a sample relative to the objective lens.  Furthermore, a good mechanical stage is one of the key distinguishing features of research-grade microscopes compared to the many low cost microscopes currently available\cite{Breslauer:2009,*Cybulski:2014,*Bogoch:2013,*Tapley:2013,*Lee:2014}.  Thus, our design has the potential to enable a wide range of experimemts that are impossible with current low-performance microscopes but are difficult due to the significant size and cost of current research microscopes -- timelapse experiments requiring days or weeks of microscope time, use in constrained environments such as fume hoods or incubators, or applications in containment labs requiring disposable equipment.  We have used the Raspberry Pi single-board computer and camera module\cite{RasPiWebsite} as its small size and low power consumption make it suitable to run automated experiments on the microscope for days or weeks.

\fig{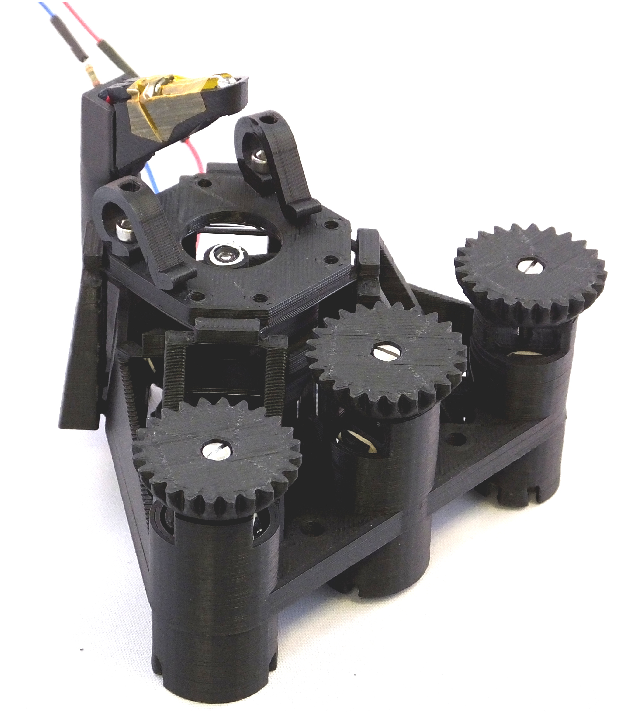}{Photograph of the microscope.  The three gears at the front control lateral motion (outer gears) and focus (centre), while the sample is held on the translating stage by two printed clips.  A white LED is mounted on a printed arm at the top, and the lens (from the Raspberry Pi camera module) is visible through the hole in the sample stage.  The camera sensor mounts underneath the microscope.}

\section{3D Printed Hinges}
The explosion of interest in 3D printing in recent years has extended to science; the technique has been exploited by an increasing number of researchers to address cutting-edge research needs through custom chemical reactionware\cite{Symes:2012}, open-source optomechanical components\cite{Zhang:2013} and vortex chambers\cite{Geertsen:2015} among many examples.  

Flexure hinges, also referred to as ``living hinges'', operate through deformation of their constituent material\cite{Paros:1965}.  Specific points in the structure are deliberately weakened by making them thin, so that the hinge bends reversibly under stress.  Such mechanisms are typically metal, but the greater compliance of plastic compared to metal allows a longer range of motion in flexure joints\cite{Trease:2005}, and can thus be an advantage.  However, the mechanism must be carefully designed to account for plastic's lower stiffness compared to metal.  Our aspiration is that flexure-based moving parts will allow not only microscopes but a great variety of useful mechanisms to be printed for scientific\cite{Pearce:2014} and other applications.

While high-end professional 3D printers can work with multiple materials, including some specifically designed to flex, our microscope is intended to be printed on the basic RepRap-type machine.  Such machines form their 3D structures by depositing a filament of molten plastic onto the part being printed, building it up layer by layer.  Overhanging parts beyond about $45^\circ$ must therefore be supported during printing, which requires time-consuming removal of support material (more sophisticated machines often use a different material for support that can be dissolved away).  No support material is required to build our microscope, as we have avoided cantilevered parts.  The most challenging parts to print are the thin flexures (the precise dimensions of which may need to be adjusted for different printers and materials), and the various ``bridges'' where a span of plastic on an upper layer joins two disconnected regions.  On a correctly-adjusted machine this is not a problem, but we have included a test object in the design to enable the printer to be optimised without printing a whole microscope.

If the structure is deformed beyond its elastic limit, permanent damage will result and the lifetime of the mechanism will be short.  In our design, the range of motion is intentionally limited to avoid bending any flexure hinge through more than  $\alpha = 6^\circ$ (see \rfig{flexure_geometry}).  We print the stage such that the hinge axes all lie in the horizontal plane: this allows us to make the flexures thinner and stronger than if the hinge axis is vertical, as the printer's layer height is thinner than the minimum width of the extruded plastic in the $xy$ plane.  

Corner filleted flexure hinges\cite{Lobontiu:2001} are used rather than the circular cut-outs normally machined from metal, as this is a better match with the layer-by-layer fabrication method of a 3D printer.  After a number of iterations of the design, the optimised flexure links are $t=0.72\,$mm thick, $l=1.5\,$mm long, and $4\,$mm wide.  Assuming the plastic bends into an arc (\rfig{flexure_geometry}) we can estimate the strain in the top and bottom of the flexure link.  The radius of curvature will be $\nicefrac{l}{\alpha}$ and thus the maximum strain  $\Delta=\nicefrac{t\alpha}{2l}\approx 0.024$.  This is close to the yield strain of both the PLA and ABS plastics usually employed in 3D printing\cite{Tymrak:2014}.

\fig{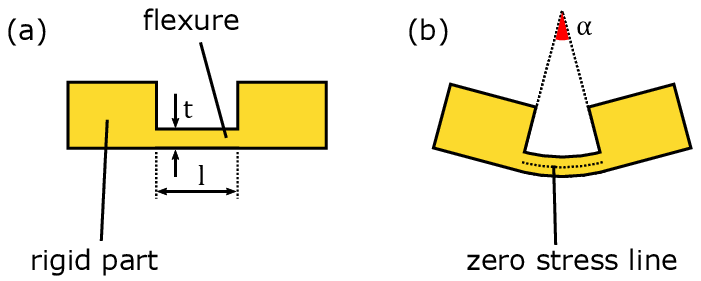}{Geometry of a flexure link, (a) relaxed and (b) bent.  Approximating the shape of the deformed flexure as an arc allows estimation of the stress experienced by the plastic.}

The maximum strain is only reached at the very edge of the flexure, hence we believe the plastic deforms in a  reversible manner.  This is borne out in practice as none of our microscope stages have yet failed due to flexures snapping, even after 6 months of use.  In designs where the motion of the flexures is less well constrained, however, we have observed flexures snapping due to being bent beyond safe limits.  The parameters $l$ and $t$ can be easily set in the parametric CAD design for the microscope, simplifying any adjustments needed if it is to be printed in a material requiring different flexure geometry.

\section{Mechanical design}
In a high powered microscope, it is important to be able to accurately focus and position the sample.  Doing so by hand requires a deal of patience and practice, and keeping the microscope in place over the course of an experiment without a suitable mount is impossible.  Our translation stage uses flexure hinges connected by rigid links.  This forms a system of levers (\rfig{pin_jointed_structures}), so that the table-shaped structure on which the sample is mounted can move in a two-dimensional plane.  The objective is then mounted on a four bar mechanism, which can move up and down to focus the microscope.  All three axes are actuated by M3 screws and nuts, which can be controlled using printed thumb-wheels or printed gears meshing with compact stepper motors.

\fig{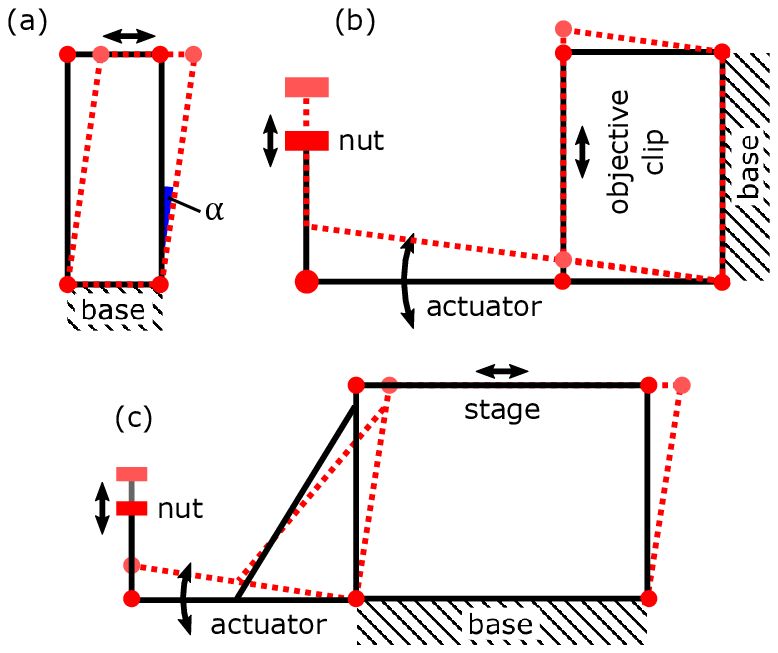}{The microscope's mechanism, represented as 2D pin-jointed structures: (a) a basic 4-bar mechanism, allowing the top part to translate in one dimension, (b) the $z$ axis mechanism, and (c) the $x$ or $y$ mechanism.  Deformed positions are shown with dashed lines, and the angle through which hinges are bent, $\alpha$, is shown in (a).}

\fig{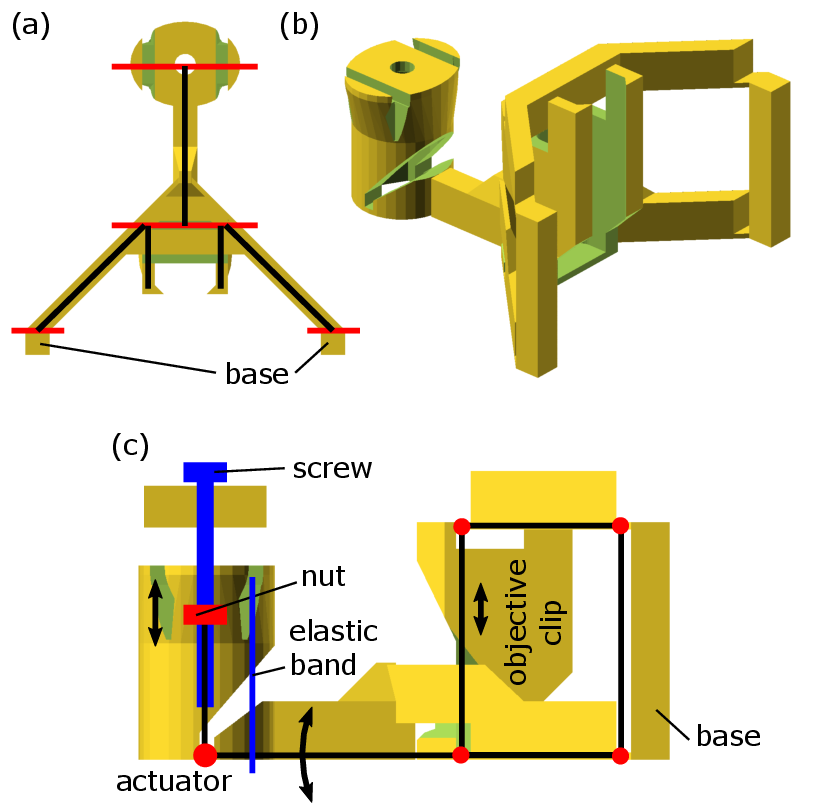}{Flexure mechanism for vertical motion of the microscope objective, (a) plan view, (b) orthographic projection, and (c) elevation showing the flexure hinge points as circles. }

Parallelogram structures form the basis of the microscope mechanism: the four bar linkage in \rfig{z_mechanism} allows the objective to translate vertically without changing orientation or lateral position.  As the flexible parts of the structure bend, the path of the moving part is an arc, but over the range of motion we use it is very close to linear. The total range of motion is limited by the maximum angle through which the flexible part of the structure can be bent, here $\alpha\approx6^\circ$.  This gives a usable range of around $20\%$ of the lever length (i.e. $\pm10\%$), while the maximum deviation from a straight-line path is only $1\%$ of the lever length.  In our case, the defocus observed at the extremes of the $xy$ travel of the stage is easily compensated for by adjusting the focus.  

\fig{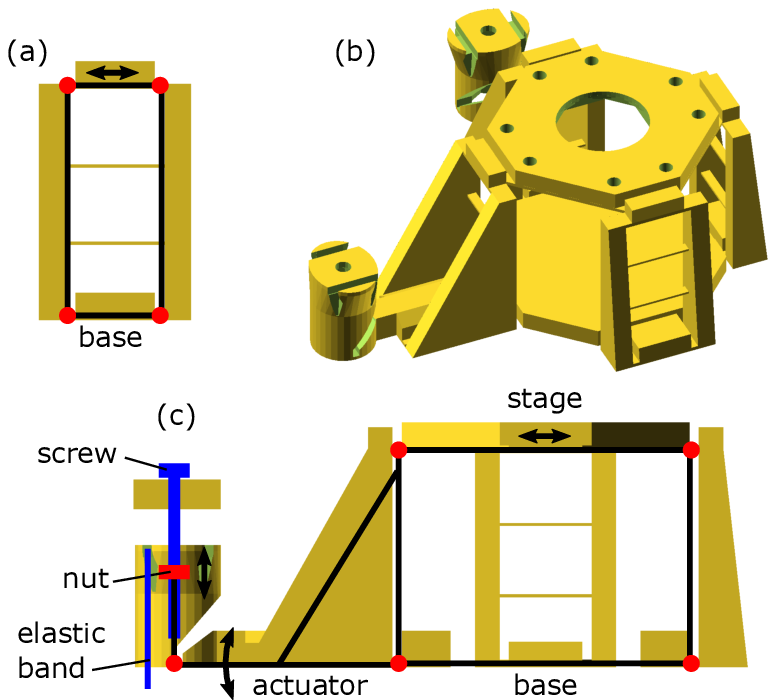}{(a) Each leg is a parallelogram, allowing it to flex and accommodate motion in one direction. (b) Orthographic view of the $xy$ subassembly. (c) Another parallelogram is formed by leg, actuator, and stage, where the actuator transmits motion from the screw to the stage. }

Motion in two degrees of freedom is often achieved by stacking two stages at $90^\circ$ to each other.  This is effective in very stiff metal designs\cite{Gao:1999}, but would result in a very large planar structure for the range of motion we require.  Such a structure, printed in plastic, would have unacceptably low stiffness out of the plane (i.e. in the focus direction).  We instead employ a table-like structure  (\rfig{stage_mechanism}). This arrangement is stiffer and more compact than stacked stages, as there is a very direct link between the $xy$ stage and the microscope body.  Our design also results in much lower bending moments on the rigid parts of the structure.  The actuators are connected to the stage via levers, each of which moves only in $x$ or $y$, but the legs of the table-like structure can tilt to allow motion in both $x$ and $y$. This means that both actuators are static - they are rigidly mounted to the microscope body and vibrations are not as  readily transmitted to the stage.

The pitch of a standard M3 screw is $0.5\,$mm per revolution, and with the addition of a thumbwheel it is possible to achieve motion with sub-micron precision by hand.  Using widely-available Arduino microcontrollers\cite{ArduinoWebsite} and RAMPS electronics\cite{Jones:2011}, we use stepper motors (with 2:1 reduction by gears) and $\nicefrac{1}{16}$-step control to achieve approximately $50\,$nm microsteps (see  \rfig{steps}).  In the case of the $x$ and $y$ axes, motion of the M3 nut is transferred to the stage with a ratio just below unity but the $z$ axis lever is designed to mechanically reduce the motion by a factor of $2.6$ to improve precision.

The whole microscope body prints as a single piece, which is responsible for the high stiffness and low drift of the structure.  However, in order to print reliably and without support material, it is necessary to avoid cantilevered structures and very thin vertical parts.  This is the reason for the use of eight legs rather than four; it allows the legs to be linked together with bridges, and those bridges to be linked together to form the stage.  This avoids the need for any part to be compliant along two axes (which would necessitate it being very thin\cite{Lobontiu:2003a}) while also allowing the stage to be printed without support.  All the moving parts are supported by the print bed during printing, and are then freed when the microscope is removed at the end.

Printed feet are added to the bottom of the microscope to allow the actuating levers to protrude below the bottom of the structure.  The feet do not impact the microscope's stability as the important mechanical linkages between the sample and the objective and (to a lesser extent) the objective and the sensor, do not depend on the feet.  The lens holder clips in afterwards in order to provide coarse adjustment, using a dovetail clip that exploits the layered structure of the material to lock it in place and prevent slipping.  In total, there are ten parts clipped on to the microscope body after printing (four feet, three gears to actuate the screws, and holders for the illumination, camera, and objective lens).  Printing the main structure takes around eight hours on our RepRapPro Ormerod, and around five on the commercial machines we have tested (Ultimaker 2 and MakerBot 2).  It uses approximately 90 grams of plastic. 

\section{Mechanical Performance}
Relative separation of the stage and the objective is the quantity that must remain constant for stable imaging.  This was measured by imaging $6\,\mu$m latex spheres attached to a glass slide, then performing particle tracking using cross-correlation with a reference image.  For most measurements, this was done in real time using Python on the Raspberry Pi.  For high speed measurements, camera frames were recorded to RAM at $90\,$Hz then written to disk and analysed offline.  These particle tracking experiments allowed us to quantify both mechanical drift and the accuracty and repeatability of the stage when driven by stepper motors.  Drift in the $z$ axis was measured by folding the imaging path with a prism, such that the objective lens was turned through $90^\circ$.  The $x$ axis on the camera became the $z$ axis, and the microscope slide was mounted on printed supports to stand it up vertically.

\subsection{Drift}
The table-shaped mechanism supporting the $xy$ stage ensures that, while it can be moved in $x$ and $y$ by adjusting the actuator screws, it is held rigidly in the third axis ($z$).  Owing to the small size and light weight of the microscope, the mechanical linkage between the sample stage and the objective lens is stiff and exhibits remarkably low drift; this is ideal for timelapse experiments and helps to minimise the influence of external vibrations on the microscope.  To quantify the drift, we tracked the motion of the sample over periods of days or weeks. Over a period of 5 days, the focus $z$ (the axial distance between objective lens and sample) typically drifted by less than $10\,\mu$m, as illustrated in \rfig{drift}.  The $x$ and $y$ axes exhibit similar drift, though they can suffer more from creep, i.e. slow plastic deformation of the structure, due to the higher forces involved due to the greater number of flexure joints.  This can result in greater drift if the stage has been moved by a large amount immediately prior to an experiment, as seen in the right hand graph of \rfig{drift} where the $y$ trace appears to relax exponentially at the start of the experiment.

\fig{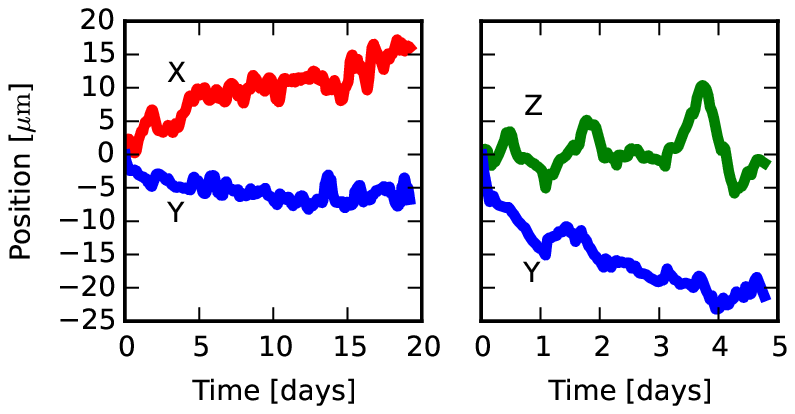}{Drift in the stage position as a function of time, left in a non air-conditioned room for several days.  The position of the stage was measured by tracking a $6\,\mu$m latex bead stuck to a microscope slide. Lateral position was measured over two weeks, while axial position was tracked over one week using a turning mirror to rotate the objective lens $90^\circ$}

The $z$ axis has fewer flexure joints and thus requires less force to move, resulting in lower creep and faster decay of stress caused by movement.  Lateral drift is less detrimental to most timelapse experiments as it can be corrected by analysis software\cite{Schneider:2012,Thevenaz:1998} provided the objects of interest do not drift out of the field of view.  The axial drift is low enough that the microscope does not lose focus. This may need to be improved if a higher numerical aperture lens is used, but is more than adequate for the lens used here.

Allan Deviation is used as a metric of stability on different timescales.  Here, it is calculated by dividing the time series $x(t)$ into chunks of length $\tau$, and calculating the mean position of each chunk $$\bar{x}_i =\frac{1}{\tau} \int_{t=i\tau}^{(i+1)\tau} x(t) \mathrm{d}t.$$The Allan deviation is the mean squared difference between adjacent chunks:
$$\mathrm{Allan\ Deviation} = \sqrt{\left\langle \left(\frac{\bar{x}_{i+1}-\bar{x}_i}{2}\right)^2 \right\rangle}.$$
This measures the mean drift over a given time period $\tau$, and has been used to assess optical tweezers systems for stability \cite{Gibson:2008,*Czerwinski:2009}.  The Allan Deviation of our system is shown in \rfig{allan_deviation}.  While our system does not attain the extremely low noise floor of these highly optimised systems, it is nonetheless within an order of magnitude, representing adequate performance for a great many experiments. This few-nm performance is astonishing from such a simple device.

\fig{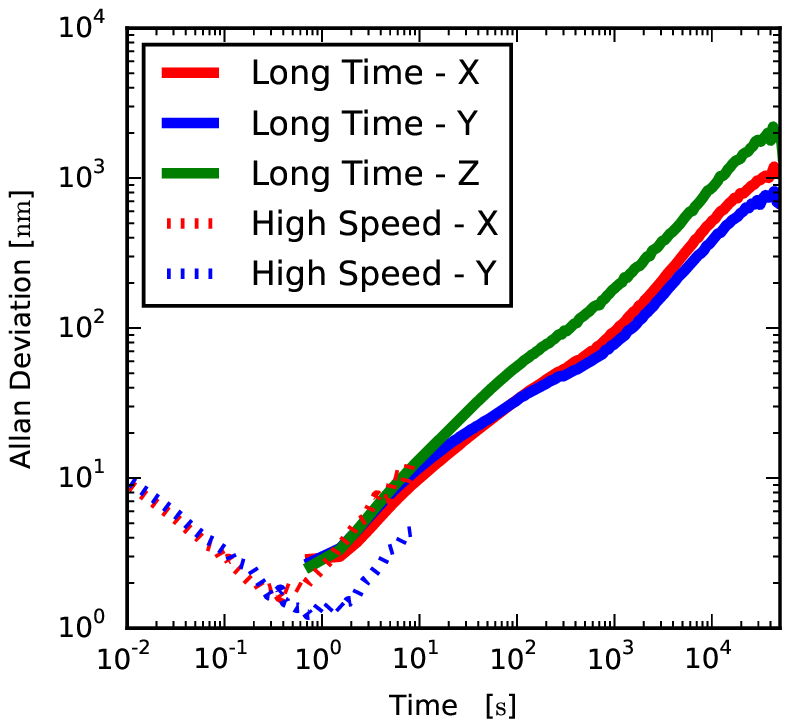}{Allan Deviation of the relative position of stage and objective, calculated from the time series data in \rfig{drift}.  The Allan Deviation on short timescales was calculated based on a shorter dataset acquired at $90\,$Hz (``high speed''), shown as dotted lines on the graph.  The downward-sloping region at small times is caused by measurement noise, while the upward slope at longer times indicates drift.}

On short timescales, the Allan deviation decreases as $\tau^{-1/2}$, corresponding to averaging over measurement noise.  On longer timescales the deviation increases again due to drift.  Here, the slope is between $\tau^{1/2}$ (diffusive motion) and $\tau$ (linear drift), suggesting that the drift is not simply diffusive but that it is also not simply linear creep.  High speed measurements show that the lowest variance between time chunks occurs around one second, which is where the trade-off between measurement noise and instrument stability has a minimum.

\subsection{Accuracy \& Repeatability}
An important parameter in any positioning device is how accurately it can return to a given position after moving away.  To assess this, the automated stage fitted with stepper motors was moved by a random distance and direction, and then moved back (see \rfig{repeatability}).  Repeatability was less than $1\,\mu$m for small moves, increasing to $\sim 15\,\mu$m for moves of $1\,$mm.  Our metric for repeatability is the root mean squared error when moving from one point to another a given distance away then returning to the start point. Performance is improved by the use of an anti-backlash algorithm, which always approaches the target point from the same direction.

\fig{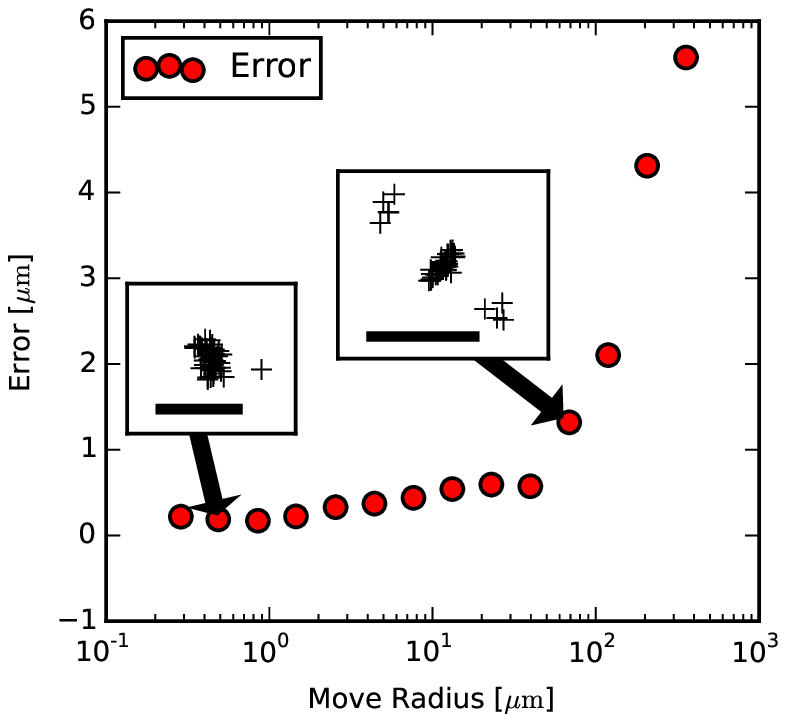}{RMS positioning error after moving a set distance in a random direction and returning to the origin.  Insets show scatterplots of the points the stage returned to for moves of two distances.  Scale bars indicate $1\,\mu$m and $4\,\mu$m in the smaller and larger insets respectively.}

Measurement of the distance travelled by the stage as a function of step size reveals that the stage moves slightly less than expected. While a simple calculation based on thread pitch and mechanical reduction suggests $250\,\mu$m per revolution or $78\,$nm per microstep of a $2:1$ geared motor, the actual distance travelled is $67\pm6\,$nm, about $86\%$ of this.  We attribute the discrepancy to flex in the actuating lever, which appears to be repeatable as the material remains within its elastic limit.  \rfig{steps} shows the stage response to moving the motor in increments of single microsteps.

\fig{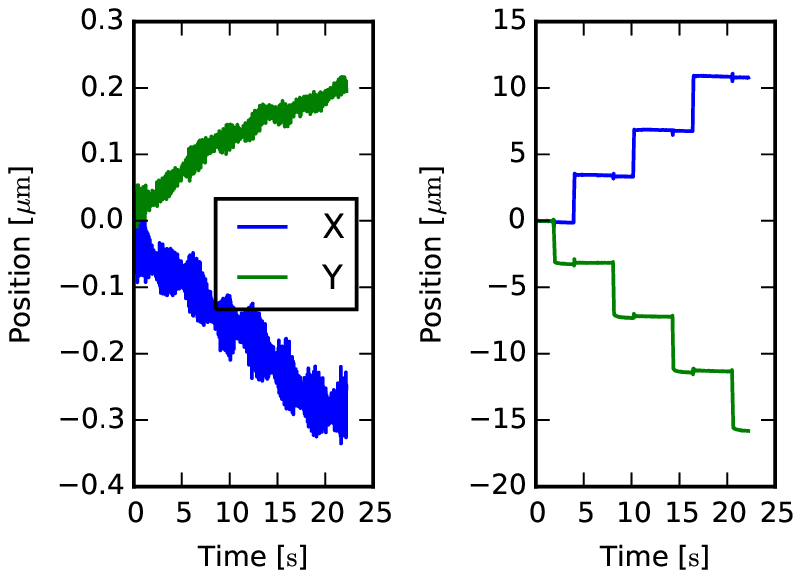}{The stage response to making steps alternatively in X and Y.  The size of single microsteps (left) size is small, but vibration caused by pulse width modulation in the motor driver is visible on small length scales. Larger steps (right) are better defined.}

\section{Optics}
There are a variety of ways to obtain the optics for a reasonable-quality microscope extremely simply. Here, we used the lens from the Raspberry Pi camera module.  This high-performance mobile phone lens is designed to focus light from infinity onto the $1.4\,\mu$m pixels of the sensor, and so has a short focal length\cite{Kinch:2015} ($3.6\,$mm) and a relatively low f number of $2.9$.  We reverse the lens, so the side designed to be next to the sensor faces the sample, and mount it just over one focal length from the sample (\rfig{pollia_and_optics}(c)).  Placing the sensor $47\,$mm from the lens magnifies the sample so that each pixel corresponds to $120\,$nm  in the sample, and the field of view is $300\times230\,\mu$m.  The optical resolution (around $2\,\mu$m) and field of view is comparable to a good modern 10x or 20x microscope objective with a numerical aperture of $0.15$, though $120\,$nm per pixel is more reminiscent of the magnification obtained with a $50\times$ or $100\times$ objective.  The camera module can bin pixels together, enabling higher-speed, lower-noise imaging at up to $90\,Hz$.  As the pixels are significantly smaller than the optical resolution of the system, this does not lose valuable information.

The basic configuration of our microscope uses transmission illumination: this gives bright-field images, suitable for observing many transparent samples.  Adding a condenser lens in a printed holder allows dark-field and basic phase contrast imaging, allowing a greater range of samples to be observed (\rfig{pollia_and_optics}).  The inverted design of the microscope means that samples are generally imaged through slides or coverslips.  Inverted microscopes work well with a wide range of samples, including cell cultures and microtomed specimens.  As our primary aim has been to develop the mechanical platform, there are many opportunities for improvement that we intend to pursue, for example adding fluorescence imaging, using a better low-cost lens\cite{Sung:2015}, or enhancing the resolution\cite{Tian:2015}.
     
\fig{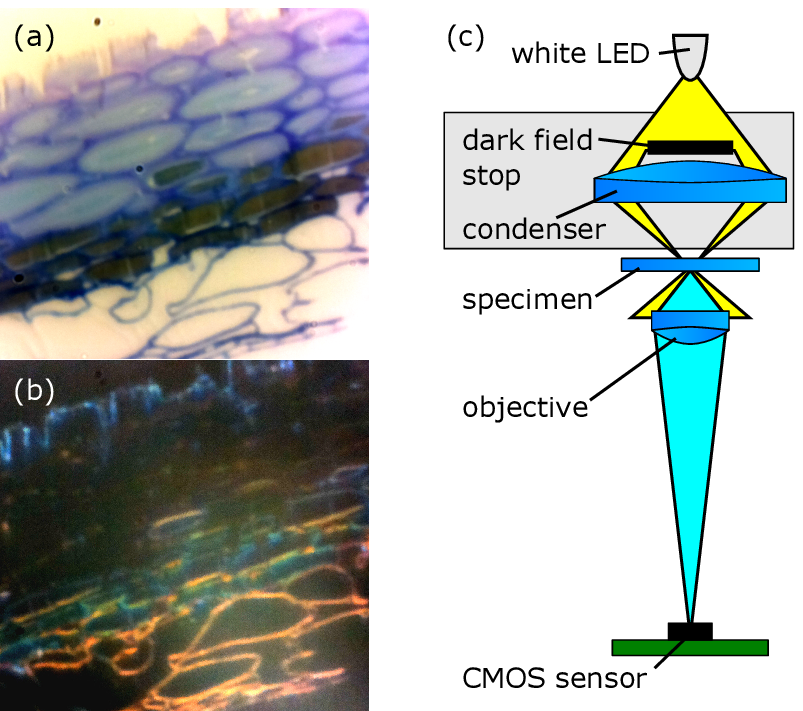}{A microtomed section of \textit{Pollia condensata} fruit\cite{Vignolini:2012} imaged in (a) bright field and (b) dark field modes.  (c) The imaging optics in the microscope, showing optional condenser lens and dark field stop.  Removing the dark field stop converts the microscope to bright field mode, and removing the condenser lens decreases the brightness but does not prevent the microscope from working.}

\section{Conclusions}
We have presented a high-performance translation stage that can be manufactured by 3D printing.  By exploiting the flexibility of the plastic materials used, we have achieved a range of motion that is greater than that available from metal flexure stages, together with sub-micron position accuracy.  The monolithic design also exhibits remarkably low drift even in ambient conditions over days or weeks.  This translation stage can form the basis of a high-performance microscope based on readily-available optics, with sufficient mechanical stability to perform timelapse experiments without autofocus or active drift correction.  As well as being a useful characterisation tool for the translation stage, this microscope is a useful tool in its own right for timelapse experiments and applications where space and weight is at a premium.  Open-source design files and assembly instructions are freely available\cite{openflexure_microscope_docubricks} and can be printed on the vast majority of currently-available machines.  We hope our design enables custom translation stages to be integrated into the growing library of open-source hardware; there is much potential for other 3D printed flexure-based mechanisms, and we intend to further investigate such applications in the future.

\section{Acknowledgements}
We would like to thank Paula Rudall (Jodrell Laboratory, Royal Botanic Gardens, Kew, UK) for preparing the \textit{Pollia condensata} samples.  RWB was supported by Research Fellowships from Queens' College, Cambridge and the Royal Commission for the Exhibition of 1851, and partial support was provided by EPSRC EP/L027151/1, the University Teaching and Learning Innovation Fund and the SynBioFund initiative.

Data supporting this publication is available at \url{http://www.repository.cam.ac.uk/handle/1810/253294}.  Design files and assembly instructions are available at \url{http://docubricks.com/projects/openflexure-microscope}.

\bibliography{library}

\end{document}